\pdfoutput=1
\documentclass[sigconf]{acmart}

\AtBeginDocument{%
  }

\copyrightyear{2026}
\acmYear{2026}
\setcopyright{cc}
\setcctype{by}
\acmConference[UIST '26]{The 39th Annual ACM Symposium on User Interface Software and Technology}{November 02--05, 2026}{Detroit, MI, USA}
\acmBooktitle{The 39th Annual ACM Symposium on User Interface Software and Technology (UIST '26), November 02--05, 2026, Detroit, MI, USA}
\acmDOI{10.1145/3830398.3830670}
\acmISBN{979-8-4007-2856-3/2026/11}

\usepackage{comment}
\usepackage{graphicx, makecell, array}
\usepackage{tabularx, booktabs, ragged2e} 
\usepackage{multirow} 
\usepackage{enumitem} 
\usepackage{subcaption} 

\begin{document}

\title[DOBI]{DOBI: Dynamic Opportunistic Body Input via Spare Joint Recruitment for Hands-Free XR}

\author{Rachel Kim}
\orcid{0000-0002-3428-4967}
\affiliation{\department{School of Computing}\institution{KAIST}\city{Daejeon}\country{Republic of Korea}}
\email{rachel02@kaist.ac.kr}

\author{Xun Qian}
\authornote{Sang Ho Yoon and Xun Qian are Co-corresponding authors.}
\orcid{0000-0003-1976-7992}
\affiliation{\institution{Google}\city{San Jose}\state{California}\country{USA}}
\email{me@xun-qian.com}

\author{Sang Ho Yoon}
\authornotemark[1]
\orcid{0000-0002-3780-5350}
\affiliation{\department{Graduate School of Culture Technology}\department{School of Computing}\institution{KAIST}\city{Daejeon}\country{Republic of Korea}}
\email{sangho@kaist.ac.kr}

\renewcommand{\shortauthors}{Kim et al.}

\begin{abstract}
Extended Reality~(XR) systems are often most useful when users are engaged in ongoing physical tasks, yet current interaction techniques still largely assume the hands are available. 
We present opportunistic body input, an interaction paradigm that redirects continuous XR control to whichever available body region remains free in the moment. 
To investigate how users naturally coordinate these spare-body movements, we conducted an elicitation study across six hand-busy scenarios. 
We found that while users' preferred spare body regions shift dynamically based on physical constraints, the resulting spontaneous movements share a consistent, low-dimensional kinematic structure organized around a dominant principal axis. 
Building on these findings, we present DOBI~(Dynamic Opportunistic Body Input), a real-time XR technique that uses gaze to target a UI element, a brief trigger gesture to identify the recruited spare body region, and the region's subsequent motion to drive continuous 1D control.
A 1D Fitts' law study establishes the baseline motor performance of this paradigm across four distinct body regions, achieving throughputs up to 2.62 bits/s with an overall 5.0\% error rate, and a dual-task usability study shows that DOBI supports reliable, low-effort control (SUS~=~84.2) while users remain engaged in realistic hand-busy activities.
\end{abstract}



\begin{teaserfigure}
  \includegraphics[width=0.95\textwidth]{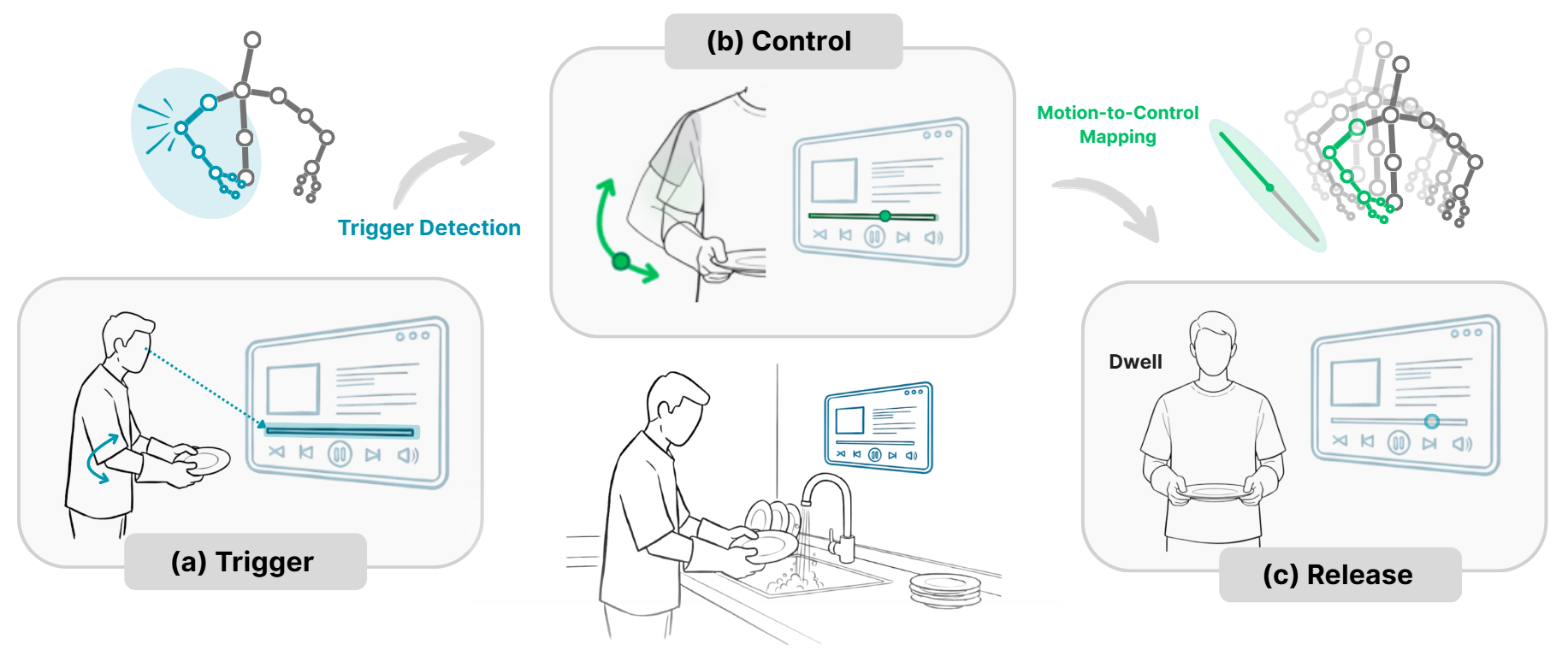}
  \caption{Overview of Dynamic Opportunistic Body Input~(DOBI) in a hand-busy context. While washing dishes, the user (a) looks at a UI target and performs a brief trigger gesture with an available spare-body movement (e.g., elbow flock), (b) controls the UI element through motion along the region's principal axis, and (c) ends the interaction with steady-state motion~(e.g., holding still).}
  \Description{Line-drawn illustration of a person washing dishes at a sink while a floating media-player window hovers beside them. Three labelled panels surround the central scene. Panel (a), Trigger: the person holds a plate in both hands, a dotted gaze ray runs from their eyes to the player's progress bar, and a blue arrow marks a short back-and-forth elbow movement; an accompanying skeleton diagram highlights the recruited left-arm branch in blue. Panel (b), Control: a close-up of the bent elbow with green arrows showing motion along a single axis, and the player's progress handle sliding along the bar in response; a second skeleton diagram, labelled Motion-to-Control Mapping, shows the same limb chain in green with a green ellipse marking its dominant principal axis. Panel (c), Release: the person holds still with the plate in both hands, labelled Dwell, and the player's handle has come to rest at its new position.}
  \label{fig:teaser}
\end{teaserfigure}

\maketitle

\section{Introduction}

Extended reality~(XR) promises a future in which digital content can seamlessly be integrated into everyday activities. 
To realize this vision, the way users control and manipulate UI elements, such as adjusting sliders, scrolling, or rotating a 3D object, should remain situated and seamless, preserving the continuity of our ongoing tasks.
In practice, however, most current XR interactions assume that the hands are always available, relying on mid-air hand gestures such as raycasting, pinching, and direct manipulation~\cite{monteiro2021handsfree}.
While effective in controlled settings, these techniques often become awkward and fatiguing~\cite{hincapie2014consumed} in the very moments when XR could be most useful, particularly when the hands are already occupied by an ongoing physical task.

To support hands-free UI manipulation, prior work has explored many alternatives to hand input, including voice~\cite{harada2006vocal, funk2020non}, gaze~\cite{ramirez2021gazehold, rolff2025hands}, EMG~\cite{saponas2009enabling}, and facial expressions~\cite{jingu2023lipio, lai2024gazepuffer}. Together, these efforts demonstrate that meaningful interaction can be performed without the use of hands. However, each modality constitutes a single, fixed input channel and therefore inherits its own contextual limitations. For instance, voice input may be socially inappropriate in public settings~\cite{easwaramoorthy2014voice} and degrades significantly in noisy environments~\cite{clark2019speech}; gaze is effective for pointing but suffers from the Midas touch problem~\cite{jacob1993eye} and induces fatigue during continuous control~\cite{hirzle2020eye, majaranta2014eye}; EMG signals are sensitive to electrode shift and require per-session calibration, limiting practical robustness~\cite{eddy2023emg}; and facial expression techniques are confined to a small, discrete command vocabulary~\cite{costanza2007intimate} that can be socially conspicuous in public~\cite{profita2013dont}.

To address these limitations, the body's skeletal movement offers a particularly rich resource. The body comprises many independently movable joints~(shoulders, elbows, wrists, knees, ankles, or the head)~\cite{wagner2013body, chen2012extending}, so that when some are engaged by a physical task, others inevitably remain free. People already exploit this capacity informally, tapping a phone with an elbow when hands are wet or nudging a door with a shoulder while carrying boxes.

Prior systems have successfully established the feasibility of leveraging body regions as expressive input sources, including the wrist~\cite{buschek2018extending, fashimpaur2023investigating, youn2022wristmenu}, forearm~\cite{harrison2010skinput, weigel2014more, hamdan2017run}, torso~\cite{van2024gestureshirt, karrer2011pinstripe}, and feet~\cite{muller2019mind, tsai2024gait, muller2023tictactoes}. Collectively, these works provide foundational evidence that the body is a viable interaction medium. However, these techniques were predominantly designed as static, single-location interfaces that permanently anchor controls to a predefined anatomical site. If that region becomes occupied by a physical task, the interaction channel is lost. Adaptive frameworks such as Human I/O~\cite{liu2024humanio} have recognized that available input channels shift with situational context, and BodyScape~\cite{wagner2013body} has mapped the body's interaction design space across regions. Yet no existing system provides a runtime mechanism for accepting an arbitrary spare-body movement, identifying the active joint, and deriving a control mapping on the fly.

These observations motivate an opportunistic framing of a novel body input technique in XR. Prior work in tangible interaction has described opportunistic controls as interfaces that leverage the physical affordances of the environment~\cite{henderson2008opportunistic}. 
We apply the same principle to the body itself. Here, we use opportunistic body input to map interactions to the spontaneous movements of spare body regions, rather than restricting them to a fixed body region.

Realizing this concept of opportunistic body input requires empirical grounding.
It remains an open question (1) what movements users naturally choose when physically constrained, and (2) whether the resulting articulations share sufficient kinematic regularity to support a general-purpose control mapping.

To explore this design space, we conducted an elicitation study in which participants physically enacted six hand-busy scenarios and improvised ad-hoc articulations to achieve seamless and continuous control.
Our analysis suggests that while preferred body regions vary across contexts, sampled control poses within a chosen articulation are often well approximated by a dominant first principal component.
Building on these observations, we present DOBI~(Dynamic Opportunistic Body Input), a real-time XR interaction technique that operationalizes opportunistic body input~(Figure~\ref{fig:teaser}).
DOBI combines gaze targeting with a brief trigger gesture to identify the recruited body region. Then, the system translates motion along the movement's principal axis directly into continuous control of the UI element.
Through a controlled 1D Fitts' law study and a dual-task usability study, we demonstrate that DOBI affords effective continuous control, establishing the baseline motor performance across four motions and showing that it remains usable and low-effort while users are engaged in realistic hand-busy activities.

In summary, we make the following contributions:
\begin{enumerate}
    \item \textbf{Opportunistic Body Input.} An interaction paradigm that redirects continuous UI control to whichever body region is free in the moment.

    \item \textbf{Empirical Characterization of Spare-Body Motion.} A formal analysis of spontaneously generated movements, identifying a consistent, low-dimensional kinematic structure in pose space that persists across diverse body regions and physical contexts.

    \item \textbf{The DOBI Technique.} A real-time XR interaction technique utilizing a trigger-conditioned pipeline to map spare body movements into continuous 1D UI control signals.

    \item \textbf{Baseline Performance \& Usability.} A 1D Fitts' law study establishing the baseline motor performance (throughput and linearity) across four distinct body regions, and a dual-task usability study demonstrating that DOBI remains reliable and low-effort under realistic hand-busy conditions.
\end{enumerate}

\section{Related Work}

\subsection{Hands-Free Interaction in XR}

When hands or visual attention are occupied, prior work has
turned to a range of alternative input channels: voice and
non-verbal audio~\cite{mihara2005migratory, sporka2006non,
zielasko2015blowclick, funk2020non, zielasko2017reliable},
gaze~\cite{adams2008inspection, ramirez2021gazehold,
rolff2025hands}, facial or oral
gestures~\cite{jingu2023lipio, lai2024gazepuffer,
wilson2025interface}, and muscle activity sensed through
EMG~\cite{shen2022clenchclick, eddy2023framework}. These
techniques demonstrate that meaningful hands-free
interaction is achievable, each offering a distinct
combination of discrete activation and continuous
control, with pointing treated as a special case of the latter.

However, each modality is tied to a single fixed channel
and therefore inherits contextual limitations: voice is
unreliable in noisy or socially sensitive
settings~\cite{rico2010usable, koelle2020social}, gaze is
effective for pointing but fatiguing for sustained
continuous control~\cite{jacob1990what, majaranta2014eye},
and facial or EMG-based signals offer compact but
specialized input. Adaptive systems such as Human
I/O~\cite{liu2024humanio} have made this fragility explicit
by cataloging which channels remain available across
everyday situations. Our work departs from this
single-channel paradigm by treating the body as a pool of
channels from which users recruit dynamically based on
current availability.

\subsection{Body Movement as Input}

While prior work has explored the body as a touchable 
surface, such as acoustic sensing on the 
arm~\cite{harrison2010skinput}, textile controls in 
clothing~\cite{karrer2011pinstripe}, and touch on the 
ear~\cite{kikuchi2017eartouch}, cheek~\cite{yamashita2017cheekinput}, 
or lap~\cite{mi2023laptouch}, these systems typically 
still require a free hand to perform input and therefore do not directly address hand-busy interaction.

More relevant work treats body \emph{movement} as 
the input signal. Prior systems have used the wrist and forearm 
motion for scrolling and shortcut 
control~\cite{fashimpaur2023investigating, 
buschek2018extending}, torso movement for eyes-free 
gestures~\cite{van2024gestureshirt}, and lower-limb input 
such as foot taps~\cite{muller2019mind}, toe 
flexion~\cite{muller2023tictactoes}, and gait 
gestures~\cite{tsai2024gait}. Frameworks such as 
BodyScape~\cite{wagner2013body} further characterize how 
different body regions can support interaction. 
Collectively, these works confirm that the body offers a 
rich input space beyond the hands.

These systems each demonstrate that a specific body
region can serve as an effective input channel. Our work
asks a complementary question: rather than optimizing
interaction for one predetermined region, can a single
system support whichever region happens to be available?
This requires both a mechanism for dynamic region
recruitment and a control mapping that generalizes across regions: extracting axis and range at runtime from the user's own motion, and assigning polarity implicitly from movement direction, rather than specifying any of these at design time.

\subsection{Opportunistic and Context-Adaptive Interaction}

Our conceptual framing draws on research in opportunistic 
and context-aware interaction, where systems adapt to the 
resources available in the current situation. In tangible 
and ubiquitous computing, prior work has repurposed nearby 
objects and surfaces as spontaneous 
controls~\cite{corsten2013instant, 
henderson2008opportunistic, hettiarachchi2016annexing, 
he2023ubi, he2024adaptui}, demonstrating that interaction 
need not be bound to predefined physical interfaces. In XR,
adaptive systems have responded to changing modality
availability~\cite{liu2024humanio} and layout or content
optimization based on the user's situation~\cite{li2024situationadapt},
as well as ergonomic constraints such as comfort, fatigue, or
grasp state~\cite{evangelista2021xrgonomics,
cheng2022comfortable, hincapie2014consumed,
cheng2023interactionadapt, aponte2024grav,
zhou2020gripmarks}.

These systems adapt interaction to external context---nearby objects, social situation, or the user's ergonomic
state. Human I/O~\cite{liu2024humanio} extends this to input
itself, but at the level of which whole channel (e.g., voice,
gaze, body) is available. We extend this logic one level
further, to the body's own internal availability: rather than
adapting \emph{what} is displayed, \emph{where} an interface
is placed, or \emph{which channel} is active, we adapt
\emph{which region within the body} is active, treating spare
upper-body regions as opportunistic resources that can be
recruited and released as the user's task demands change.

\section{Design Elicitation Study}

The concept of opportunistic body input assumes that users can intuitively repurpose spare limbs to control continuous values. 
To explore this, we conducted a design elicitation study to examine which spare-body movements users naturally produce during hand-busy tasks, and whether these elicited articulations share a universal kinematic structure for continuous control.

\begin{figure}[t]
  \centering
  \includegraphics[width=0.83\columnwidth]{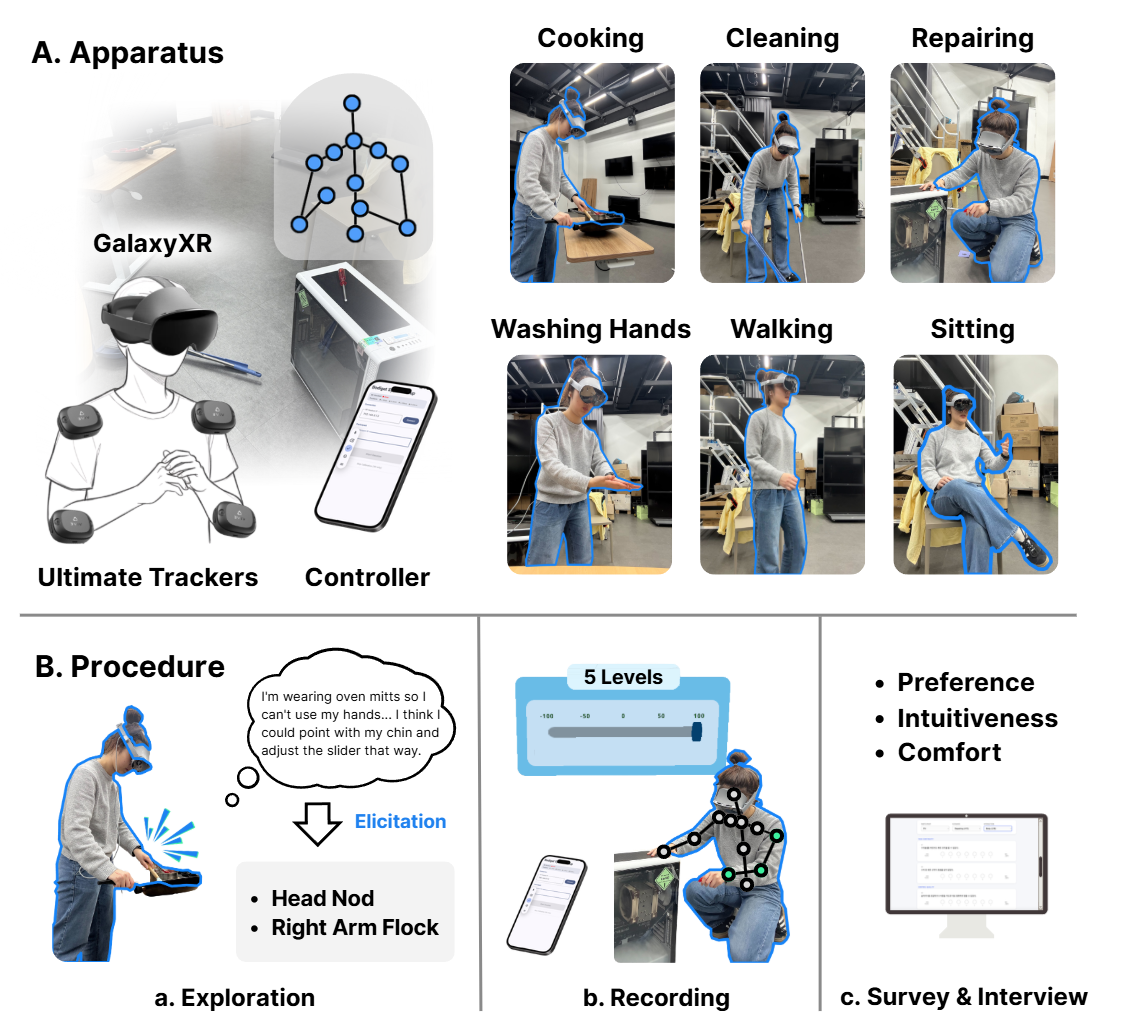}
  \caption{Design elicitation study. (A) Tracking apparatus and six everyday hands-busy scenarios. (B) Procedure involving a. think-aloud exploration, b. kinematic recording, and c. survey and interview.}
  \Description{Two-part study figure. Part A, Apparatus, shows a Galaxy XR head-mounted display, a set of body-worn Ultimate Trackers strapped to the upper arms, forearms and torso of an illustrated wearer, a handheld controller with a phone-based survey app, and an inset node-link diagram of the tracked upper-body skeleton. To its right, six photographs show a participant wearing the headset in each hands-busy scenario, with the participant's silhouette outlined in blue: Cooking (stirring a pan at a table), Cleaning (pushing a floor mop), Repairing (crouching and holding a tool), Washing Hands (hands together over a basin), Walking (mid-stride), and Sitting (seated with legs crossed). Part B, Procedure, shows three sequential stages. Stage a, Exploration: a participant cooking with a thought bubble reading ``I'm wearing oven mitts so I can't use my hands... I think I could point with my chin and adjust the slider that way,'' with an arrow labelled Elicitation pointing to the elicited movements Head Nod and Right Arm Flock. Stage b, Recording: a phone beside a participant overlaid with a tracked joint skeleton, and a floating slider UI marked with five levels from minus 100 to 100. Stage c, Survey and Interview: a monitor showing a questionnaire, with the measured constructs listed as Preference, Intuitiveness, and Comfort.}
  \label{fig:elicitation-study}
\end{figure}

\subsection{Participants and Apparatus}
We recruited 12 experienced XR users (> 10 hours of HMD usage) to ideate and gain insight into spare body movements ($M=26.1, SD=4.0$; 7 male, 5 female; all right-handed).
In order to record these movements, participants wore a GalaxyXR headset and four VIVE Ultimate Trackers strapped to both shoulders and elbows, allowing full upper-body skeletal tracking without external base stations.
The experimenter operated a companion mobile application to operate the headset remotely, allowing participants to keep their hands engaged in the ongoing activity throughout the session.

\subsection{Scenarios and Task}
Hand-busy situations refer to everyday contexts where the user's hands are occupied by a physical task limiting their ability to interact with
devices~\cite{sarsenbayeva2018situational}. Drawing on prior taxonomies of situational impairments~\cite{wobbrock2019situationally, liu2024humanio}, we selected six everyday
scenarios that vary in posture, mobility, and physical constraint on the hands (Figure~\ref{fig:elicitation-study}A): \textbf{Cooking, Cleaning, Repairing, Washing Hands, Walking, and Sitting}. To ensure realistic immersion, each scenario included:

\textbf{Physical Props:} tools for given contexts (e.g., a frying pan and spatula for cooking, or a screwdriver and desktop computer for repairing)

\textbf{Contextual Applications:} real applications overlaid in the FOV, providing targets for the control (e.g., a YouTube recipe video, music player)

While engaged in the primary physical activity, participants used a think-aloud protocol to propose and execute spare-body movements for a concurrent control task, such as adjusting a virtual parameter~(e.g., a slider).

\subsection{Procedure}
Prior to the main trials, the instructor guided participants through a warm-up session. To mitigate legacy bias, participants were encouraged to explore the full range of motion of their head, shoulders, elbows, and wrists, helping them recognize various body parts as potential interaction surfaces beyond the hands.

Following this warm-up, participants completed the six scenarios in a randomized order. Each scenario followed a three-step procedure (Figure~\ref{fig:elicitation-study}B):

\textbf{Step 1: Exploration.}
Using the provided tools and applications, participants acted out the primary tasks within the scenario. Through think-aloud exploration, they identified specific ``hand-busy moments'' where they needed to interact with the system but could not use their hands. Within that context, participants ideated all spare-body movements they felt were available and natural for controlling a UI element.

\textbf{Step 2: Recording.} 
Once a movement was finalized, we recorded its kinematic trajectory. To observe the movement's directionality, polarity, and operational intervals, participants were asked to enact the articulation across five discrete intended control levels: from Level -2 (e.g., maximum decrease) to Level +2 (e.g., maximum increase), passing through Level 0 (the neutral resting posture). When the user held each pose, the experimenter remotely captured the full upper-body skeletal frame using the companion mobile application.

\textbf{Step 3: Subjective Assessment.}
Participants rated seven upper-body regions (head, bilateral shoulders, elbows, and wrists) for intuitiveness and comfort using 7-point Likert scales, followed by a semi-structured interview to capture their selection rationale.

\subsection{Findings}
In total, the elicitation study yielded 437 free-form spare-body input designs across the 6 scenarios, 426 of them with valid recordings (11 were removed due to tracking loss). Because opportunistic body input assumes that users can improvise a workable control movement on the spot, we analyzed this data to understand how users naturally recruit their bodies under hand-busy constraints, and to determine whether the resulting movements can be formalized into a general-purpose control mapping.

\begin{figure}[t]
  \centering
  \includegraphics[width=0.83\columnwidth]{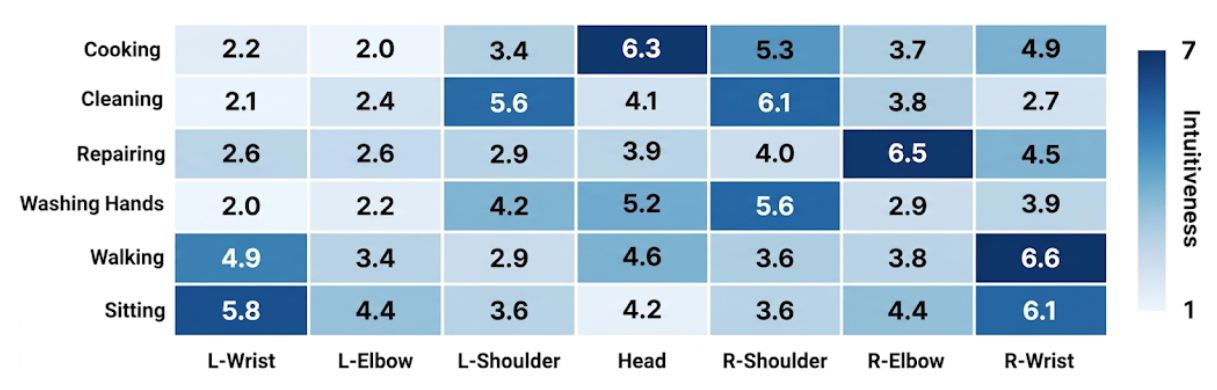}
  \caption{Mean intuitiveness ratings (7-point Likert) for seven upper-body regions across six activity scenarios.}
  \Description{A six-row by seven-column heatmap of mean intuitiveness ratings on a 1-to-7 scale, with darker blue indicating higher ratings. Rows are the activity scenarios and columns are the body regions L-Wrist, L-Elbow, L-Shoulder, Head, R-Shoulder, R-Elbow, and R-Wrist, in that order. Cooking: 2.2, 2.0, 3.4, 6.3, 5.3, 3.7, 4.9. Cleaning: 2.1, 2.4, 5.6, 4.1, 6.1, 3.8, 2.7. Repairing: 2.6, 2.6, 2.9, 3.9, 4.0, 6.5, 4.5. Washing Hands: 2.0, 2.2, 4.2, 5.2, 5.6, 2.9, 3.9. Walking: 4.9, 3.4, 2.9, 4.6, 3.6, 3.8, 6.6. Sitting: 5.8, 4.4, 3.6, 4.2, 3.6, 4.4, 6.1. The highest-rated region differs from scenario to scenario: the head for cooking, the right shoulder for cleaning and washing hands, the right elbow for repairing, and the right wrist for walking and sitting.}
  \label{fig:preference}
\end{figure}

\paragraph{\textbf{F1. Preferred input region varies by scenario.}}
Prior body-centric interaction research has established that multiple body locations can serve as input surfaces~\cite{chen2012extending, wagner2013body}, and individual systems have demonstrated viable input from the wrist~\cite{buschek2018extending}, forearm~\cite{harrison2010skinput}, torso~\cite{van2024gestureshirt}, and feet~\cite{muller2019mind}. 
However, these systems each commit to a single, fixed body region at design time.
Our results show that no single region should be fixed this way.
Intuitiveness ratings (Figure~\ref{fig:preference}) revealed that even the globally highest-rated region (right wrist, $M=4.78$) ranked first in only two of six scenarios, dropping substantially in the others.
Instead, the top-rated region changed according to the given task (Figure~\ref{fig:preference}; Friedman tests, all $p<.01$).
As P3 noted, ``Wrist rotation feels easier, but when my hands are wet, using my elbow feels more intuitive.'' 
This provides empirical motivation for dynamic region recruitment: the user, not the system, must initiate which region to use.

\paragraph{\textbf{F2. Elicited movements can be summarized by a small set of articulation families.}}
\begin{figure}[t]
  \centering
  \includegraphics[width=0.83\columnwidth]{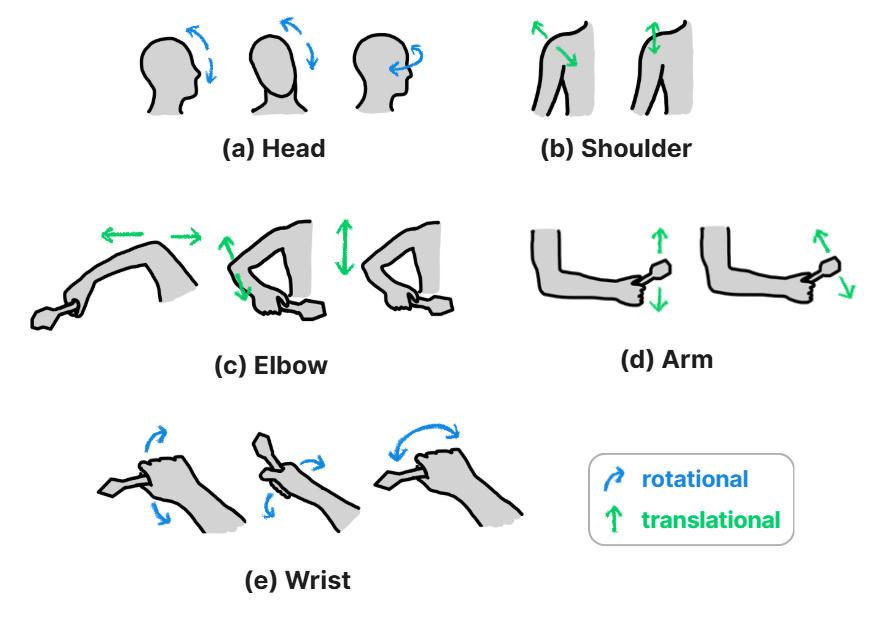}
  \caption{Representative articulation families elicited in the study across upper-body regions. Blue arrows indicate dominant rotational motion; green arrows indicate dominant translational displacement.}
  \Description{Five groups of grey line drawings of body parts, each annotated with motion arrows; a legend states that blue curved arrows denote rotational motion and green straight arrows denote translational motion. Group (a), Head: three side and front views of a head with blue curved arrows showing nodding up and down, tilting side to side, and turning left and right. Group (b), Shoulder: two views of a shoulder and upper arm with green arrows showing shrugging up and down and shifting diagonally. Group (c), Elbow: three views of a bent arm with green arrows showing the forearm extending outward and the elbow rising and falling. Group (d), Arm: two views of a bent arm held at waist height with green arrows showing the forearm moving vertically and diagonally. Group (e), Wrist: three views of a hand and forearm with blue curved arrows showing wrist flexion and extension, rotation about the forearm axis, and side-to-side sweeping.}
  \label{fig:elicited-movements}
\end{figure}

Gesture elicitation studies have shown that users converge on shared gesture sets for discrete commands such as pinch-to-select or swipe-to-scroll~\cite{pierella2015remapping}. These studies primarily focus on categorical gesture types and measure inter-participant agreement. We extend this methodology from discrete gestures to continuous, analog body movements. 

Although participants were free to devise any movement, design outputs clustered into 11 recurring articulations~(Figure~\ref{fig:elicited-movements}). Participants varied widely in \textit{which} region they recruited~(F1), but not in \textit{how} they moved a given region once selected. This convergence is not a matter of participants agreeing on a named gesture; rather, each body region's anatomy admits only a small number of natural, controllable motions (e.g., a wrist can rotate or flex, but has little else to offer as a continuous 1D signal), so independent participants converge on the same few articulations by default. The practical implication is that the system does not require open-ended motion recognition: once a body region is recruited, its movement can be drawn from this small, predictable set of anatomically favored articulations.

\paragraph{\textbf{F3. User-designed movements are kinematically well approximated by a dominant 1D axis.}}

During think-aloud sessions, we observed that users expressed control intent in two distinct ways. When using distal joints (head and wrist), participants described rotating the joint in place (e.g., P11: \textit{``I'll twist my wrist to turn the volume up''}). When using proximal joints (elbow and shoulder), participants instead focused on displacing the limb through space (e.g., P4: \textit{``I'll drag my elbow like a pointer sideways''})---even though this apparent translation is still produced by rotating a single joint further up the arm.

\begin{figure}[t]
  \centering
  \includegraphics[width=0.664\columnwidth]{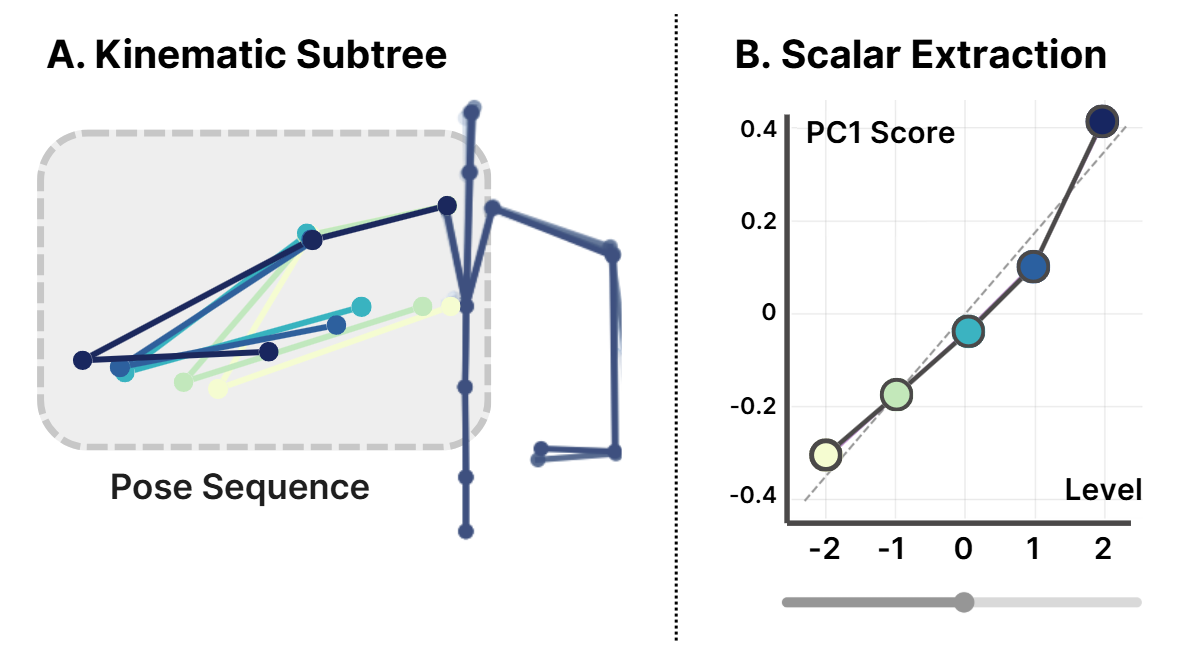}
  \caption{Example of Pose Linearization. (A) Pose vectors are extracted exclusively from the recruited kinematic subtree across five intended control levels (from -2 to +2). (B) Projecting these multi-joint poses onto their first principal component (PC1) reveals a highly linear 1D trajectory, confirming the movement can be mapped directly to a continuous control signal.}
  \Description{Two side-by-side panels. Panel A, Kinematic Subtree, shows a stick-figure skeleton drawn in dark blue; the recruited arm branch is enclosed in a dashed grey rounded rectangle labelled Pose Sequence, and five overlaid copies of that branch, coloured from pale yellow-green through teal to dark navy, fan out to represent the five intended control levels. Panel B, Scalar Extraction, is a line plot with the control Level from minus 2 to 2 on the horizontal axis and PC1 Score from minus 0.4 to 0.4 on the vertical axis. Five circular markers, coloured to match the poses in Panel A, sit at approximately minus 0.31, minus 0.17, minus 0.04, 0.10, and 0.41 and lie almost exactly on a dashed straight reference line, showing a near-linear relationship. A horizontal slider control is drawn beneath the plot.}
  \label{fig:kinematic-analysis}
\end{figure}

To capture both strategies in a single representation, we model each joint's state as a 6D vector combining its 3D relative rotation (via logarithmic map) and 3D position delta: $\mathbf{v}_J = [\mathbf{r}_J, \mathbf{d}_J]^T \in \mathbb{R}^6$. For a recruited kinematic subtree $S$ of $K$ joints, these are concatenated into a unified pose vector $\mathbf{x} \in \mathbb{R}^{6K}$ (Figure~\ref{fig:kinematic-analysis}A).

Applying PCA to these pose vectors confirmed that the elicited movements are kinematically well-suited for linear 1D extraction, though we note that sampling only five discrete intended levels per articulation constrains how finely this linearity ($R^2$) can be estimated, and treat this as a limitation of the current analysis (Section~\ref{sec:limitations}).
Beyond concentrating variance on a single component ($\eta_1$: median 97.6\%, IQR: 95.0--98.9\%), the movements follow straight trajectories in pose space (median $S = 0.95$) and maintain strict proportionality between intended and projected control levels (median $R^2 = 0.98$; Figure~\ref{fig:kinematic-analysis}B).
Together, these three properties (dimensionality, straightness, and linearity) indicate that users naturally default to simple, monotonic articulations when designing continuous control, and that a single linear projection over the kinematic subtree is sufficient to extract a usable control signal. This stability is not limited to the discrete elicitation setting: the same PCA-based axis extraction also produces proportional, continuous control in our real-time Fitts' law evaluation ($R^2 = 0.74$--$0.97$; Section~\ref{sec:fitts-results}), indicating that the extracted axis remains reliable under continuous, freely-timed motion.

This approach builds on the established use of PCA for extracting low-dimensional control spaces from body signals~\cite{casadio2012body, pierella2015remapping}.
However, whereas prior work requires a dedicated calibration session to fit a per-user, per-region mapping, our data shows that the 1D structure emerges spontaneously from the user's self-chosen movement.
A single component captures nearly all variance across the body regions without region-specific modeling, suggesting that the control axis could be extracted from a short motion sample alone.

Decomposing the PC1 loading vector into per-joint contributions reveals that rotation universally drives the control axis, but for multi-joint chains (elbow and shoulder), downstream joints still contribute meaningfully rather than negligibly, suggesting that including the full subtree may improve signal fidelity compared to tracking only the root joint.

\paragraph{\textbf{F4. Movement range is volatile; polarity is consistent.}}
While the control axis structure is stable (F3), the articulation range is not. Wrist movements produced the largest median PC1 span (1.09), more than double that of head movements (0.45), with substantial variation across users and contexts. As P5 noted: \textit{``When I'm walking outside, I don't want to wave my arm around, a small wrist twist is enough.''} In contrast, \textit{polarity agreement}---whether users agreed on which direction of movement should increase versus decrease the control value---was near-universal (Head 94\%, Shoulder 97\%, Elbow 95\%, and Wrist 92\%), indicating shared mental models for directional mapping. This motivates adaptive gain with implicit polarity assignment.

Prior body-based systems typically calibrate both range and polarity in a dedicated setup phase~\cite{casadio2012body}; based on this contrast, DOBI treats the two properties differently: gain is calibrated dynamically per-activation to track the volatile articulation range, while polarity is assigned implicitly from movement direction, without a separate calibration step.

Taken together, these findings suggest a specific strategy for opportunistic body input. Region choice should remain user-initiated because preferred input regions shift across contexts (F1). Once a region is chosen, the resulting movement can often be parameterized from a short motion sample using a dominant 1D axis over the recruited kinematic subtree (F3). Because movement range varies more than directional mapping across users and contexts (F4), per-activation calibration should focus on gain while preserving a consistent polarity convention. These observations motivate DOBI's system design goals.

\subsection{Design Goals}

The elicitation study suggests two complementary requirements for DOBI: flexibility in which body region is used, and consistency in how a recruited movement can be parameterized. Together with the need to avoid accidental activation during ongoing activity, these findings motivate the following design goals.

\paragraph{\textbf{DG1. Enable on-the-fly body region recruitment.}}
F1 showed that the most suitable region shifts with task constraints that are difficult for a system to predict. The interaction should therefore allow users to recruit whichever region they find available, rather than being bound to a predetermined one.

\paragraph{\textbf{DG2. Require a deliberate activation trigger.}}
Because opportunistic input coexists with ongoing physical activity, it is highly vulnerable to false positives~(the Midas Touch problem). The system requires a clear and deliberate activation trigger.

\paragraph{\textbf{DG3. Extract the control axis directly from the user's motion.}}
F2 showed that users converge on a small set of shared articulations, and F3 confirmed that each can be reduced to a single linear control dimension via PCA over the full kinematic subtree. The system can therefore extract the dominant axis from the user's motion and further generalize across body regions without joint-specific modeling.

\paragraph{\textbf{DG4. Calibrate gain dynamically while implicitly assigning polarity.}}
As F4 showed, the movement range is volatile, so the system normalizes gain based on the amplitude observed during activation. Movement polarity, by contrast, is assigned implicitly from the direction of the user's motion, since users exhibit near-universal agreement on directional mapping.

\paragraph{\textbf{DG5. Provide spatially congruent, body-centric feedback.}}
Unlike mid-air gestures that demand visual attention to a UI, body input leverages proprioception for eyes-free control. However, since the axis is extracted dynamically (DG3), the system must provide body-anchored feedback that instantly visualizes the active region and indicates which physical direction increases the control value.

\section{The DOBI System}

\begin{figure*}[h]
  \includegraphics[width=0.9\textwidth]{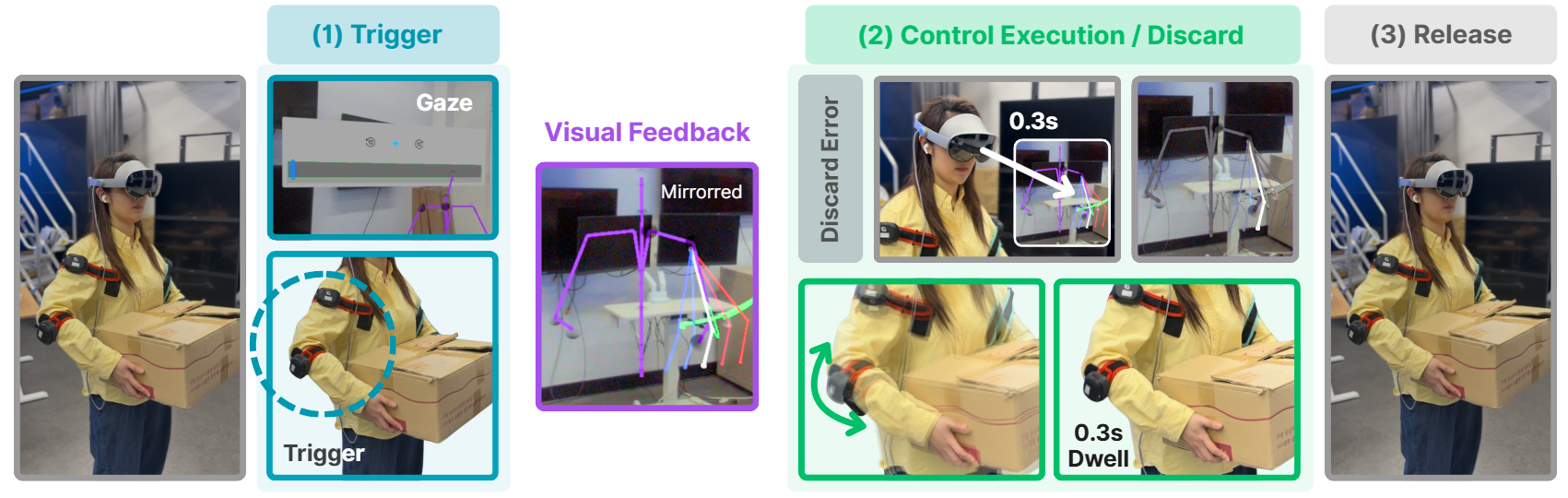}
  \caption{System Walkthrough (1) Trigger: The system detects a deliberate double-repetition gesture from a spare body region while the hands are occupied. (2) Control Execution or Discard: A mirrored skeleton visualizes the recruited limb and its motion range. The user can either manipulate the UI value through motion or discard an unintended trigger via dwell feedback. (3) Release: The interaction is committed and automatically disengaged through a 0.3s dwell, ensuring seamless task continuity without explicit exit commands.}
  \Description{A filmstrip of photographs showing a user in a lab wearing a head-mounted display and body-worn trackers while carrying a cardboard box in both hands, arranged into three labelled stages. Stage (1) Trigger, in blue: a first-person view shows the user's gaze cursor resting on a horizontal slider UI with a magenta tracked skeleton visible below it; a third-person photograph shows the user's right upper arm circled with a dashed blue ring, labelled Trigger, as it performs a short repeated movement. A separate purple-bordered inset, labelled Visual Feedback, shows the mirrored tracked skeleton rendered in the headset in magenta, blue, green and white. Stage (2) Control Execution or Discard, in green: the upper row, labelled Discard Error, shows the user gazing away and a 0.3 second timer that cancels an unintended trigger; the lower row shows the user's elbow moving along a green arc to drive the UI value, followed by a frame labelled 0.3s Dwell in which the arm is held still. Stage (3) Release, in grey: the user stands holding the box with both hands and the interaction has ended.}
  \label{fig:system-walkthrough}
\end{figure*}

Based on our design goals, we implemented DOBI (Dynamic Opportunistic Body Input), a real-time XR interaction technique that turns spare-body movements into continuous control signals. 
The interaction unfolds in three primary stages (Figure~\ref{fig:system-walkthrough}): 

\textbf{(1) Trigger:} To initiate interaction, the user looks at a target UI element and performs a deliberate double-repetition trigger gesture with their chosen spare-body movement. 

\textbf{(2) Control Execution or Discard:} Upon successful detection, DOBI instantly displays a mirrored skeleton in the user's FOV, highlighting the recruited limb and visualizing the corresponding motion and range (DG5). The user can move the joint along this axis to continuously manipulate the UI value, or, if the trigger was accidental, they can discard the interaction by dwelling at the skeleton for 0.3s. 

\textbf{(3) Commit \& Release:} Once the user reaches the desired UI value, they simply stop moving. Dwelling at the target posture for 0.3s commits the final value and automatically disengages the control, allowing the user to resume their hand-busy task without an explicit exit command.

\begin{figure*}[h]
  \includegraphics[width=0.92\textwidth]{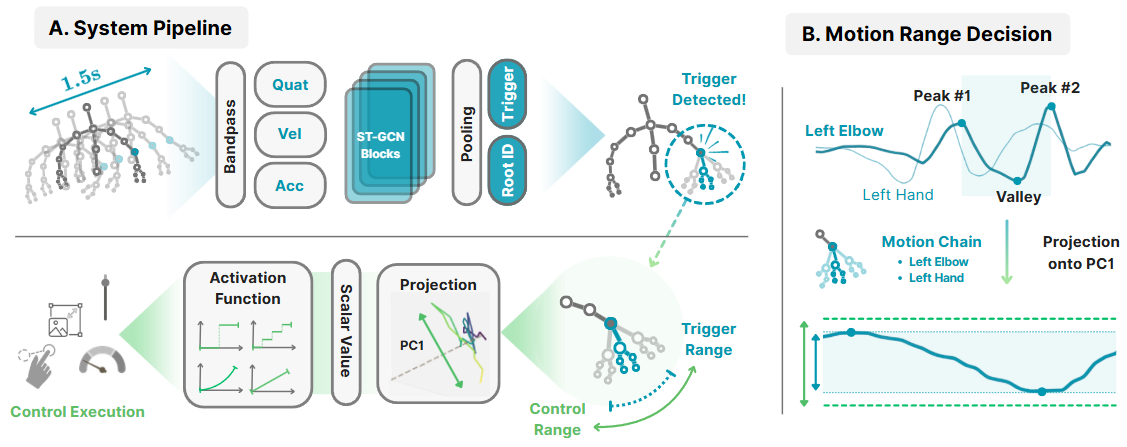}
  \caption{The DOBI interaction pipeline. (A) An ST-GCN continuously monitors a 1.5s window of upper-body skeletal data to detect a trigger and identify the active root joint. The system then extracts the kinematic subtree, projecting its motion onto PC1 to drive continuous UI activation functions. (B) Dynamic gain calibration. By analyzing temporal peaks within the trigger window and projecting the isolated motion chain onto PC1, the system automatically extracts the user's intended spatial control range..}
  \Description{Two panels. Panel A, System Pipeline, is a left-to-right block diagram in two rows. The top row starts with a skeleton sequence spanning a 1.5 second window, passes through a Bandpass block, then a column of three feature blocks labelled Quat, Vel and Acc, then a stack of ST-GCN Blocks, then a Pooling block, and finally two output heads labelled Trigger and Root ID; the result is a skeleton with the detected root joint circled in blue and captioned Trigger Detected. The bottom row continues right-to-left in green: the identified kinematic subtree is isolated and its Trigger Range mapped to a Control Range, then passed through a Projection onto PC1 block, a Scalar Value block, and an Activation Function block showing four small mapping curves, ending at UI icons for a slider, an image, a toggle and a dial under the label Control Execution. Panel B, Motion Range Decision, shows a time-series plot in which a thick teal Left Elbow curve and a thin Left Hand curve rise and fall together; two local maxima are marked Peak \#1 and Peak \#2 with the local minimum between them marked Valley, and the interval between the peaks is shaded. Below it, a small skeleton labelled Motion Chain lists Left Elbow and Left Hand, and an arrow labelled Projection onto PC1 leads to a second plot where the projected signal descends and returns; horizontal dashed green and blue lines mark the extracted control range, with a vertical double-headed arrow spanning it.}
  \label{fig:system-pipeline}
\end{figure*}

For both discarding and committing, we choose a 0.3s dwell threshold adapted from prior works in gaze interaction~\cite{majaranta2002twenty, pfeuffer2017gaze+}. 
Figure~\ref{fig:system-pipeline} illustrates the backend technical pipeline powering this interaction lifecycle, from trigger detection to the actual control execution. In the following subsections, we detail this system implementation.

\subsection{Gaze and Trigger}
To start the control, the user should perform a quick trigger motion to initiate DOBI. The trigger motion serves for two purposes: 1) to inform the system of the transition from hand input to DOBI, and 2) to hint about the control motion.

\subsubsection{Trigger Design and Data Collection}
Since opportunistic spare-body movements are highly variable, the trigger mechanism must rely on a universal temporal signature that can be consistently applied across any body region. 
Furthermore, this signature must be easily distinguishable from everyday background motions to prevent false activations~(DG2).
To satisfy these requirements, we drew inspiration from the familiar `double-click' desktop metaphor and designed the trigger as a quick, double-repetition movement~(e.g., a double wrist flick or double shoulder shrug). 
This design choice explicitly prioritizes user agency, ensuring that the system only recruits a body region when a deliberate intent is signaled, thereby mitigating the Midas Touch problem in highly dynamic, hands-busy contexts.
The user performs this trigger using their intended control motion. 
By acting as a brief ``kinematic preview,'' the gesture provides the system with the spatial data required to dynamically extract the principal control axis and calibrate the input gain.

To analyze the temporal properties of this double-repetition trigger and to train a robust detection model, we collected a targeted dataset. We recruited 12 participants~(6 female, 6 male; all right-handed) who completed 5-second trials across various behavioral and social hand-busy scenarios. In total, we collected 3916 samples, comprising two types of recordings:
\begin{itemize}[leftmargin=*]
    \item \textbf{Positive Samples:} Participants performed a deliberate double-repetition trigger with a prompted movement~(Figure~\ref{fig:elicited-movements}) in a manner that felt natural for their current physical constraint.
    \item \textbf{Negative Samples:} Rather than performing prompted non-trigger motions, participants continued to enact the hand-busy task itself (e.g., cooking, washing) while producing scenario-appropriate idle movements and single-repetition pulses. Because such incidental actions can share a similar temporal structure with a genuine trigger, both the training and held-out test sets contain these cases, forcing the classifier to strictly learn the distinctive two-repetition temporal signature.
\end{itemize}

\begin{figure}
  \includegraphics[width=0.83\columnwidth]{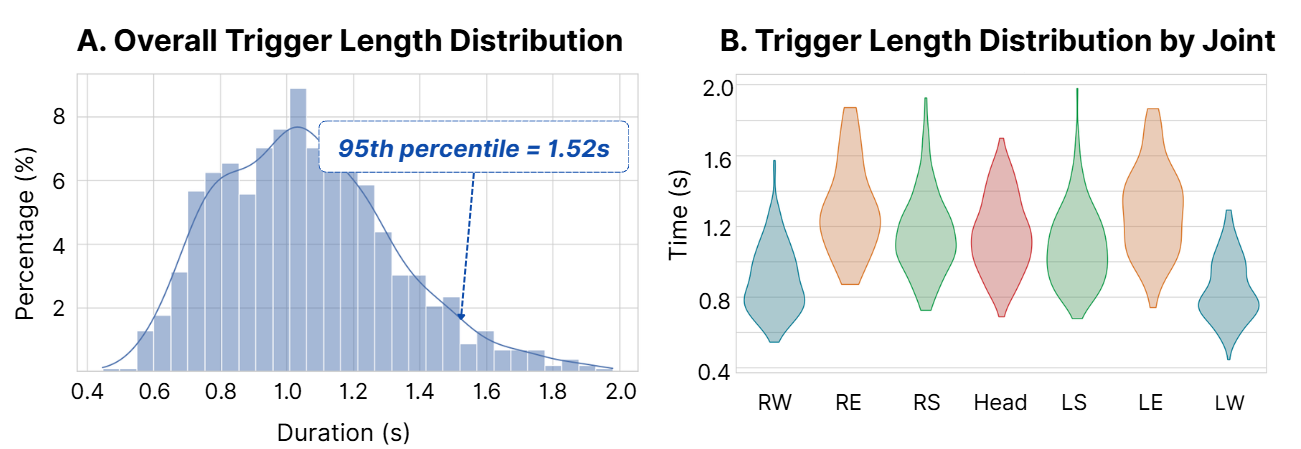}
  \caption{Trigger gesture duration: (A) overall distribution across all joints, (B) per-root joint breakdown.}
  \Description{Two panels. Panel A, Overall Trigger Length Distribution, is a histogram with an overlaid smooth density curve; the horizontal axis is trigger duration from 0.4 to 2.0 seconds and the vertical axis is percentage of trials from 0 to about 9 percent. The distribution is unimodal and right-skewed, peaking near 1.0 seconds at roughly 9 percent and tapering off past 1.6 seconds; an annotation with a dashed arrow marks the 95th percentile at 1.52 seconds. Panel B, Trigger Length Distribution by Joint, is a set of seven violin plots with duration from 0.4 to 2.0 seconds on the vertical axis, one per root joint: RW, RE, RS, Head, LS, LE, and LW. The two wrist conditions, RW and LW, are centred lowest at roughly 0.8 seconds and are the narrowest; the elbow conditions RE and LE are centred highest at roughly 1.25 seconds; RS, Head and LS fall in between at roughly 1.0 to 1.15 seconds.}
  \label{fig:trigger-analysis}
\end{figure}

\subsubsection{Trigger Characterization and Parameter Selection}

Temporal analysis of the positive samples revealed how trigger execution varies across different body parts~(Figure~\ref{fig:trigger-analysis}). Wrist triggers (RW, LW) were consistently the fastest, followed by the shoulders (RS, LS) and head, while the elbows (RE, LE) produced the slowest articulations. Despite these anatomical differences, the overall duration of the double--repetition gestures exhibited a mean of 1.06~s with a 95th percentile of 1.52~s. 
Based on this distribution, we established a fixed rolling window of 1.5~s for our real-time detection model. 
This ensures the system captures the complete double-repetition event regardless of the joint used, while filtering out excessive non-trigger context.

\subsubsection{ST-GCN Training and Evaluation}
To detect these triggers dynamically, we implemented a Spatio-Temporal Graph Convolutional Network~(ST-GCN)~\cite{yan2018spatial}. The network leverages both the spatial hierarchy of the human skeleton and the temporal dynamics of the 1.5~s rolling window. The model features two parallel output heads: a binary detection head, indicating whether a trigger is present, and a multi-class identification head, indicating which root joint drives the motion. The network was trained end-to-end using an Adam optimizer and a multi-task loss function on an NVIDIA RTX 5080 GPU.

\begin{figure}
  \includegraphics[width=0.664\columnwidth]{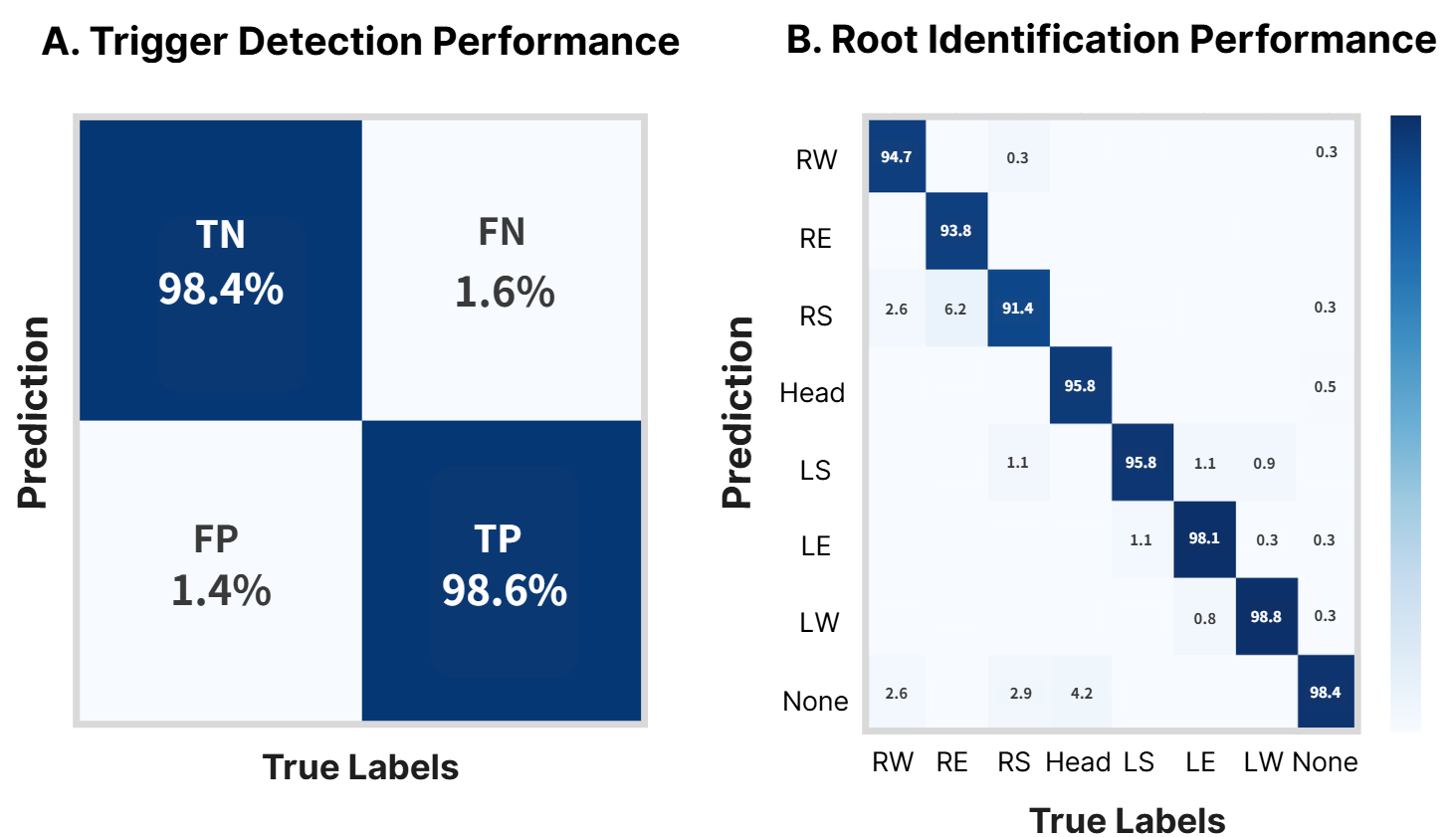}
  \caption{Quantitative performance of our model on the held-out test set. (A) The confusion matrix for the binary trigger detection task, (B) The confusion matrix for the multi-class root joint identification task.}
  \Description{Two confusion matrices rendered as blue heatmaps, with prediction on the vertical axis and true label on the horizontal axis. Panel A, Trigger Detection Performance, is a two-by-two matrix: true negatives 98.4 percent, false negatives 1.6 percent, false positives 1.4 percent, and true positives 98.6 percent, so both diagonal cells are dark and both off-diagonal cells are near white. Panel B, Root Identification Performance, is an eight-by-eight matrix over the classes RW, RE, RS, Head, LS, LE, LW, and None. The diagonal is uniformly dark, with accuracies of 94.7, 93.8, 91.4, 95.8, 95.8, 98.1, 98.8, and 98.4 percent respectively. Off-diagonal values are all below 7 percent; the largest confusions are 6.2 percent of right-elbow trials predicted as right shoulder, 4.2 percent of head trials predicted as None, and 2.6 to 2.9 percent between the right-side joints and None.}
  \label{fig:model-performance}
\end{figure}

We evaluated the architecture on a 20\% held-out test set. As shown in Figure~\ref{fig:model-performance}A, the binary detection task achieved an overall accuracy of 98.5\%. The true positive rate (recall) of 98.6\% and the exceptionally low false positive rate (1.6\%) confirm that the model is highly reliable and robust against unintended activations during hand-busy tasks. For root joint identification (Figure~\ref{fig:model-performance}B), the model achieved an average accuracy of 95.9\% across all 8 classes (7 joints + 1 none class). Rare misclassifications occurred exclusively between anatomically adjacent, functionally-linked joints (e.g., Right Shoulder vs. Right Elbow). 

Furthermore, the core ST-GCN model is designed to be lightweight, with 0.51M parameters, significantly smaller than the original ST-GCN architecture (3.1M)~\cite{yan2018spatial} and comparable to other lightweight skeleton-based models such as DD-Net (0.15M)~\cite{zhang2024dd}. Compared to other recent deep learning models deployed for XR usage (e.g., EclipseTouch~\cite{mollyn2025eclipsetouch} 3.3M, BlazePose 1.5--3.3M~\cite{bazarevsky2020blazepose}), this architecture is efficient and readily deployable on modern standalone XR headset processors.

\subsection{Motion-to-Control Mapping}
Once a trigger is successfully detected, the system must determine the precise motion axis and control range for continuous input. The ``kinematic preview'' provided during the trigger motion should be further analyzed. Figure~\ref{fig:system-pipeline}B illustrates how this brief sample is transformed into a continuous control mapping.

\subsubsection{Temporal Segmentation and Subtree Extraction}
The Trigger Detection module outputs the 1.5-second rolling window in which the trigger occurred, along with the identified root joint. To isolate the exact bounds of the user's intended motion, we perform a peak analysis within this window to identify the $P_1$ and $P_2$ peaks of the double-repetition pattern. The frames between $P_1$ and $P_2$ define the precise temporal bounds of the active trigger motion.

Using the identified root joint, we form a \textbf{kinematic subtree} containing the root and all joints distal to it in the kinematic chain (e.g., if the root is detected as the shoulder (RS), the subtree includes RS, RE, and RW). This procedure isolates the active limb and suppresses background noise from the ongoing physical task.

\subsubsection{Pose-Linearized PCA and Axis Extraction}
To dynamically extract the principal control axis, we perform pose-linearized PCA on the frames isolated between $P_1$ and $P_2$. For each joint $J$ in the subtree, we compute a 6D feature vector combining its relative rotation ($q$) and position ($p$):
\begin{equation}
    {v}_J = \begin{bmatrix} {r}_J \\ {d}_J \end{bmatrix} \in \mathbb{R}^6
\end{equation}
where ${r}_J = \mathrm{Log}(q_J \cdot q_{\mathrm{neutral}}^{-1}) \in \mathbb{R}^3$ is the log-map of the rotation relative to a neutral pose, and ${d}_J = {p}_J - {p}_{\mathrm{neutral}} \in \mathbb{R}^3$ is the position delta. The neutral pose is established as the average of the frames immediately preceding $P_1$. Concatenating all $K$ joints in the subtree yields a unified pose vector ${x} \in \mathbb{R}^{6K}$.

Applying PCA to these mean-centered pose vectors yields the first principal component ${w}_1$, which defines the 1D motion axis.

\subsubsection{Dynamic Gain Calibration and Signal Gating}
To ensure the extracted axis is suitable for continuous UI control, the system applies two quality gates before accepting the gesture:
\begin{enumerate}[leftmargin=*]
    \item \textbf{Linearity Gate:} The motion must be fundamentally one-dimensional. Using the PCA eigenvalues ($\lambda_k$), the explained variance of the first principal component must satisfy:
    \begin{equation}
        \eta_1 = \frac{\lambda_1}{\sum_k \lambda_k} \geq 0.75
    \end{equation}
    
    \item \textbf{Range Gate:} The gesture must span a sufficient physical distance to filter out unintentional micro-twitches. We project each trigger frame ${x}_i$ onto the axis by subtracting the temporal mean pose ($\mu$), yielding $p_i = ({x}_i - \mu)^\top {w}_1$, and normalize it to $\hat{p}_i \in [-1, 1]$. The total spatial span ($\rho$) must satisfy:
        \begin{equation}
            \rho = \max_i \hat{p}_i - \min_i \hat{p}_i \geq 0.15
        \end{equation}
        where the $0.15$ threshold was determined empirically through pilot testing.
\end{enumerate}

Once both gates are satisfied, DOBI latches the targeted UI element. The spatial span ($\rho$) observed during the trigger establishes the input gain. All subsequent joint motion along ${w}_1$ is projected (using the same mean subtraction $\mu$) and mapped directly into continuous scalar values.

\subsubsection{UI Activation Functions}
\label{sec:ui-mapping}
Once the motion is mapped into a normalized continuous scalar $\hat{p}_t \in [-1, 1]$, this raw 1D signal can be flexibly interpreted depending on the dimensionality and desired behavior of the target UI element. Rather than altering the core tracking pipeline, DOBI applies different activation functions to $\hat{p}_t$ to compute the final control output $u(t)$. 
By decoupling the raw kinematic extraction from the final UI activation, this same signal can be flexibly time-integrated or discretized to support rate control, step-based navigation, or binary thresholds. We demonstrate the versatility of these mapping strategies in our proof-of-concept application scenarios~(Section 6).

\section{System Evaluation}

We evaluate DOBI in two complementary studies. Section~\ref{sec:fitts-results} establishes DOBI's baseline motor performance in a controlled, single-task Fitts' law study. Section~\ref{sec:usability-study} then evaluates whether DOBI remains usable when body regions must be opportunistically recruited and switched during an ongoing physical task, in a dual-task usability study.

\subsection{Fitts' Law Study}
\label{sec:fitts-results}

\subsubsection{Study Design}
To evaluate whether DOBI's motion-to-control mapping normalizes movement into predictable, proportional 1D control, we conducted a standard 1D Fitts' Law reciprocal tapping task~\cite{bi2013ffitts, wobbrock2008error}. We recruited 14 participants (6 female, 8 male; mean age = 25.09, SD = 2.2, all right-handed), using the same GalaxyXR and VIVE Ultimate Tracker setup as Figure~\ref{fig:elicitation-study}A. Participants performed a reciprocal tapping task based on ISO 9241-9~\cite{soukoreff2004towards}, mapped to a normalized virtual slider (0.0--1.0), in a within-subjects design crossing \textbf{Input Condition} (four spare-body movements---Shoulder, Arm, Elbow, Wrist---one representative motion per category from our elicitation study, Figure~\ref{fig:elicited-movements}b--e, dominant side, Latin-square counterbalanced) and \textbf{Index of Difficulty} (five levels, Table~\ref{tab:fitts_conditions}). Each condition comprised 5 sub-blocks of 15 trials (one per difficulty level, order randomized); participants triggered the instructed movement at the start of each block and rested between sub-blocks, yielding 4200 trials total (14 $\times$ 4 $\times$ 5 $\times$ 15). We analyzed standard Fitts' Law metrics~\cite{soukoreff2004towards}: Movement Time, Error Rate, Effective Width/ID ($W_e = 4.133 \times SD_x$, $ID_e = \log_2(A_e/W_e+1)$), and Throughput ($TP=ID_e/\overline{MT}$, mean-of-means). The first trial of each sub-block was removed as warm-up, and sub-blocks with $>$50\% error were excluded (3 of 280), following standard practice.

\begin{table}[t]
\centering
\caption{Task parameters ($A$, $W$) and resulting nominal $ID$s.}
\label{tab:fitts_conditions}
\begin{tabular}{lccccc}
\toprule
 & \textbf{L1} & \textbf{L2} & \textbf{L3} & \textbf{L4} & \textbf{L5} \\
\midrule
{Amplitude ($A$)} & 0.30 & 0.45 & 0.60 & 0.70 & 0.80 \\
{Width ($W$)}     & 0.15 & 0.12 & 0.10 & 0.07 & 0.05 \\
{$ID$ (bits)}     & 1.58 & 2.25 & 2.81 & 3.46 & 4.09 \\
\bottomrule
\end{tabular}
\end{table}

\subsubsection{Model Fit and Linearity}

\begin{figure}[t]
    \centering
    \includegraphics[width=0.83\columnwidth]{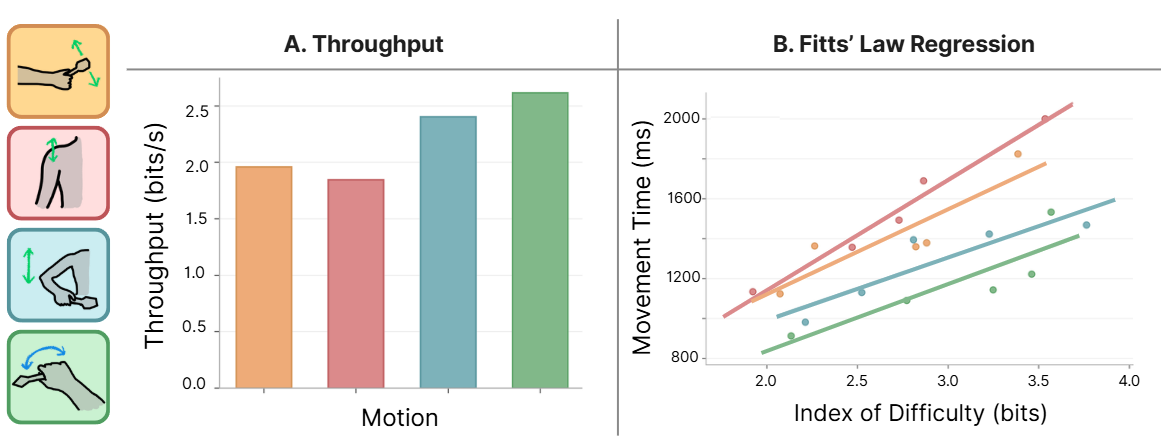}
    \caption{Fitts' Law regression models for each input condition. Each data point represents the mean $MT$ at a given $ID_e$ level, averaged across participants.}
    \Description{A composite figure. Along the left edge, four small colour-coded icons illustrate the four input conditions: an orange icon of a forearm with vertical translation arrows, a red icon of a shoulder with a vertical arrow, a teal icon of a bent elbow with a vertical arrow, and a green icon of a wrist with a curved rotation arrow. Panel A, Throughput, is a bar chart of throughput in bits per second for the four conditions in that colour order, with values of approximately 1.97, 1.85, 2.42, and 2.63; the green wrist-rotation condition is highest and the red shoulder condition lowest. Panel B, Fitts' Law Regression, is a scatter plot with regression lines, with index of difficulty in bits from about 1.8 to 4.0 on the horizontal axis and movement time in milliseconds from about 800 to 2100 on the vertical axis. Each condition contributes coloured points and a fitted line; all four lines rise linearly with index of difficulty. The red shoulder line is steepest and sits highest, the orange line lies just below it, and the teal and green lines are shallower and lie lowest, with green the fastest overall.}
    \label{fig:fitts_regression}
\end{figure}

All four body regions produced linear models with $R^2$ ranging from 0.74 to 0.97: the Shoulder was strongest ($R^2=0.97$, $p=.002$), followed by the Elbow ($R^2=0.80$, $p=.042$) and the Arm ($R^2=0.78$, $p=.047$); the Wrist was marginally significant ($R^2=0.74$, $p=.060$), likely reflecting a ceiling effect from its fast, consistent execution. These linear relationships confirm that DOBI's PCA-based mapping translates opportunistic movement into proportional 1D control consistent with standard human motor behavior.

\subsubsection{Throughput and Error Rate}

\begin{table*}[t]
\centering
\caption{Comprehensive performance and movement accuracy by input condition ($N=14$). Values are mean (SD).}
\label{tab:fitts_results}
\begin{tabular*}{0.9\textwidth}{@{\extracolsep{\fill}}lccccccc}
\toprule
& \multicolumn{3}{c}{\textbf{Performance Metrics}} & \multicolumn{4}{c}{\textbf{Accuracy Metrics}} \\
\cmidrule(r){2-4} \cmidrule(l){5-8}
\textbf{Condition} & \textbf{TP (bits/s)} & \textbf{Error (\%)} & \textbf{MT (ms)} & $W_e$ & $W_e/W$ & $A_e$ & $A_e/A$ \\
\midrule
Arm      & 1.96 (0.32) & 5.8 (4.3) & 1300 (167) & .105 (.031) & 1.03 & .492 (.180) & 0.97 \\
Shoulder & 1.85 (0.48) & 5.7 (3.4) & 1435 (296) & .099 (.030) & 0.99 & .479 (.171) & 0.93 \\
Elbow    & 2.40 (0.70) & 5.5 (3.5) & 1224 (250) & .091 (.048) & 1.01 & .539 (.194) & 0.95 \\
Wrist    & 2.62 (0.40) & 3.2 (2.7) & 1138 (163) & .082 (.025) & 0.89 & .558 (.182) & 1.02 \\
\bottomrule
\end{tabular*}
\end{table*}

Across the four regions, overall throughput averaged 2.23 bits/s (SD = 0.58) with a 5.0\% error rate (SD = 3.5\%)---comparable to prior hands-free/XR baselines ($\sim$2.47 bits/s for head pointing~\cite{hansen2018fitts}, 1.8--2.3 bits/s for gaze+pinch~\cite{wagner2023fitts}, 1.2--1.6 bits/s for foot input~\cite{velloso2015interactions}).
A repeated-measures ANOVA showed a significant effect of Input Condition on Throughput ($F(3,21)=5.84$, $p=.005$), with distal regions (Wrist, Elbow) outperforming proximal ones (Shoulder, Arm); Error Rate did not differ significantly ($F(3,21)=1.31$, $p=.30$). Effective task parameters ($A_e/A \approx 1.0$, $W_e/W \approx 0.89$--$1.03$) confirmed valid Fitts' behavior across conditions, with the Wrist showing the tightest endpoint precision (Table~\ref{tab:fitts_results}).

\subsection{Dual-Task Usability Study}
\label{sec:usability-study}

The Fitts' law study isolates DOBI's motor precision under single-task, fully instructed conditions. To evaluate whether DOBI remains usable when a body region must be opportunistically recruited and switched while a person is engaged in a realistic hand-busy activity, we conducted a dual-task usability study.

\begin{figure}[t]
  \centering
  \includegraphics[width=0.705\columnwidth]{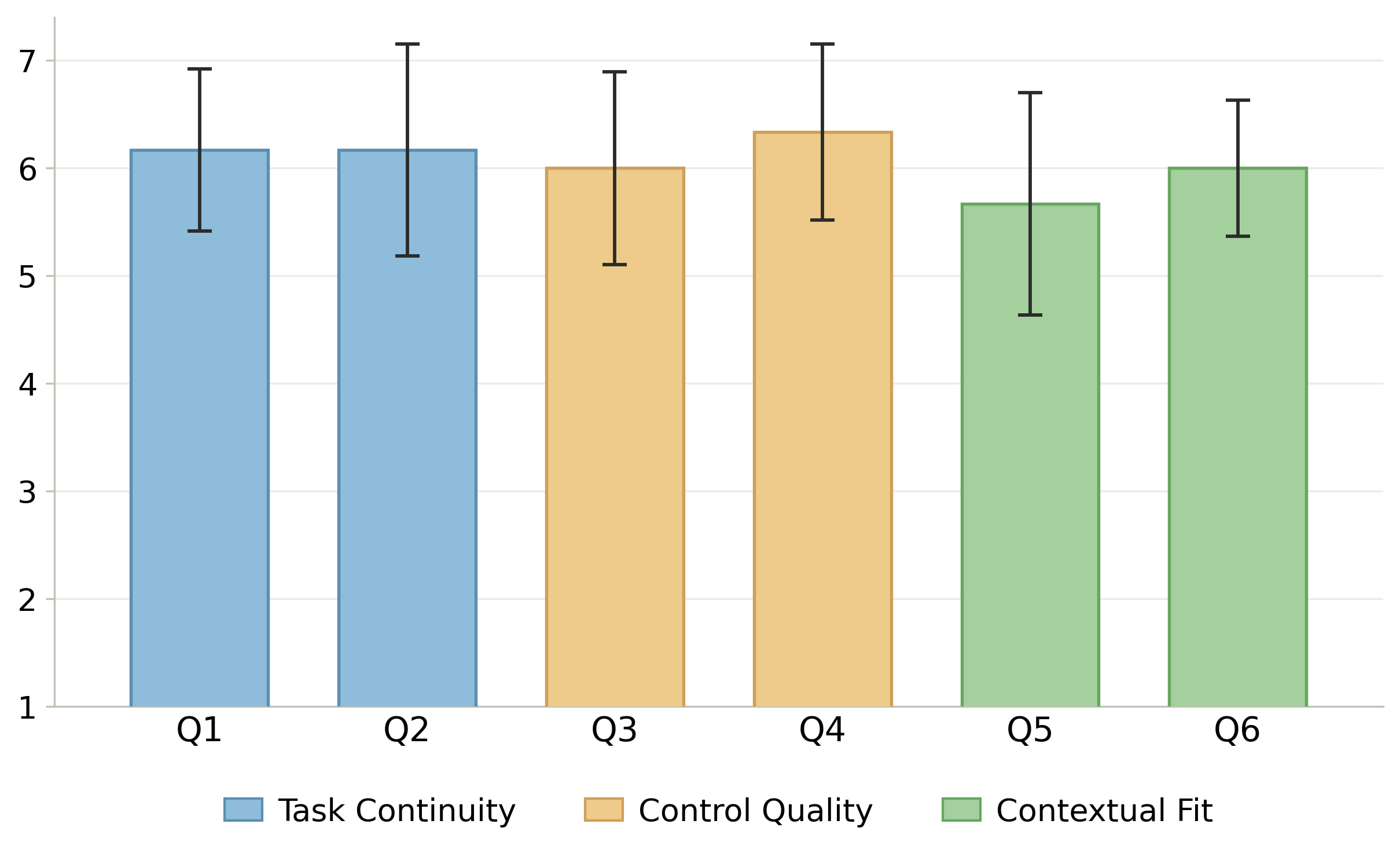}
  \caption{Per-item Likert ratings from the dual-task usability study ($N=6$). Full questionnaire items appear in Appendix~\ref{app:questionnaire}.}
  \Description{A bar chart of per-item mean Likert ratings with error bars showing standard deviation. The vertical axis runs from 1 to 7 and the horizontal axis lists six items, Q1 through Q6, colour-coded by construct: Q1 and Q2 in blue for Task Continuity, Q3 and Q4 in orange for Control Quality, and Q5 and Q6 in green for Contextual Fit. Mean ratings are approximately 6.2 for Q1, 6.2 for Q2, 6.0 for Q3, 6.3 for Q4, 5.7 for Q5, and 6.0 for Q6. All six means sit well above the scale midpoint of 4, with Q4 highest and Q5 lowest and most variable.}
  \label{fig:usability_likert}
\end{figure}

\subsubsection{Study Design}
We recruited 6 participants for a within-subjects study. Each session began with a 10-minute tutorial in which participants practiced both DOBI and a gaze-and-pinch reference condition representative of standard hands-based XR interaction. Participants then completed three primary-task scenarios drawn from everyday hand-busy activities, cooking, carrying boxes, and walking, each performed under prescribed instructions to keep task demands consistent across participants. Four interactive widgets, video, music, weather, and maps, representative of common ambient XR applications, were placed within the participant's field of view throughout each scenario.

\subsubsection{Procedure}
Participants were instructed to actively engage with the four widgets using DOBI while continuing to focus on their assigned primary task, mirroring how such interactions would occur in daily use rather than as an isolated foreground activity. Throughout each session, we logged every trigger attempt and its outcome, successful activation, false positive, or a true trigger that produced an unsatisfactory control mapping and required the user to reject and re-trigger. After completing all trials, participants completed a custom 7-point Likert questionnaire (Figure~\ref{fig:usability_likert}) and the System Usability Scale (SUS), followed by a semi-structured interview to capture qualitative feedback.

\subsubsection{Results}
Across 186 trigger attempts from the 6 participants, 173 (93.0\%) were successful, 2 (1.1\%) were false positives, and 11 (6.0\%) were true triggers with an unsatisfactory mapping that required a re-trigger. This low combined error rate indicates that DOBI's trigger detection and axis extraction remain robust when performed concurrently with an ongoing physical task, rather than only as an isolated, instructed motion.

Participants rated the system highly usable, with an average SUS score of 84.2 (SD = 5.2), well above the conventional threshold of 68 associated with above-average usability. The custom Likert ratings (Figure~\ref{fig:usability_likert}) were consistently positive across all six dimensions, highest for interaction responsiveness (Q4, $M=6.3$), ease of operation (Q1, $M=6.2$), and task flow preservation (Q2, $M=6.2$), and lowest for perceived naturalness (Q5, $M=5.7$).

\subsection{Discussion}

The Fitts' law results establish baseline motor characteristics under controlled, single-task conditions, and the dual-task study shows this capability translates into reliable, well-received interaction under realistic concurrent activity. The throughput values are a baseline rather than an upper bound: they reflect instructed motions by novice users on first exposure, and DOBI reaches this range while leaving the hands free for the primary task, unlike the comparison techniques. They also reflect the control phase only; trigger and commit overhead is discussed in Section~\ref{sec:limitations}.

The observed gradient in motor performance also provides heuristics for interaction design: distal joints (wrist, elbow) excel at fast, fine-grained continuous control, while proximal joints (shoulder, arm) are better suited to coarse adjustments or discrete actions. Accommodating this gradient does not require altering the core motion-to-control pipeline; designers can instead tailor DOBI's UI activation functions (Section~\ref{sec:ui-mapping}) to the recruited region, for example applying a 1:1 proportional mapping for precise joints or a rate-controlled function to stabilize slower, proximal movements.

Across both studies, DOBI's error sources were consistent and interpretable. In the dual-task study, failures traced primarily to individual differences in articulation style, some participants performed noticeably looser or tighter triggers than others, a source of error that per-user calibration of the trigger detector could likely reduce further.

The dual-task results also speak to how DOBI is used during ongoing activity. A participant's available body region and its range of motion change continuously and are often physically limited (e.g., while holding a box); DOBI's dynamic recruitment let participants control the UI without disrupting their posture, choosing whichever motion cost the least effort. This choice reflected both physical constraints and existing mental models (P6: ``I instinctively selected the movement with the least torque''; P2: ``zoom feels like I have to use rotation''), and P3 noted an attentional benefit of the trigger-then-release design: ``I can control without looking, after the trigger.'' Deciding which motion to use can itself add mental cost, however; rather than eliminating this cost, DOBI relocates it from a fixed, system-imposed mapping to a lightweight user choice that participants still found low-effort overall (Figure~\ref{fig:usability_likert}). This points to a design space for adaptive interfaces that suggest or pre-register suitable regions ahead of time to reduce it further.

Finally, DOBI adds to the growing space of non-hand input in XR. Alongside gaze, voice, and foot input, it offers a within-body, dynamically relocatable channel: rather than committing to one fixed modality, the choice of channel itself shifts with whatever the task leaves free. This opens further research into context-aware arbitration among non-hand channels when several are available at once.

\section{Illustrative Application Scenarios}
\label{sec:applications}

\begin{figure}[t]
    \centering
    \includegraphics[width=0.664\columnwidth]{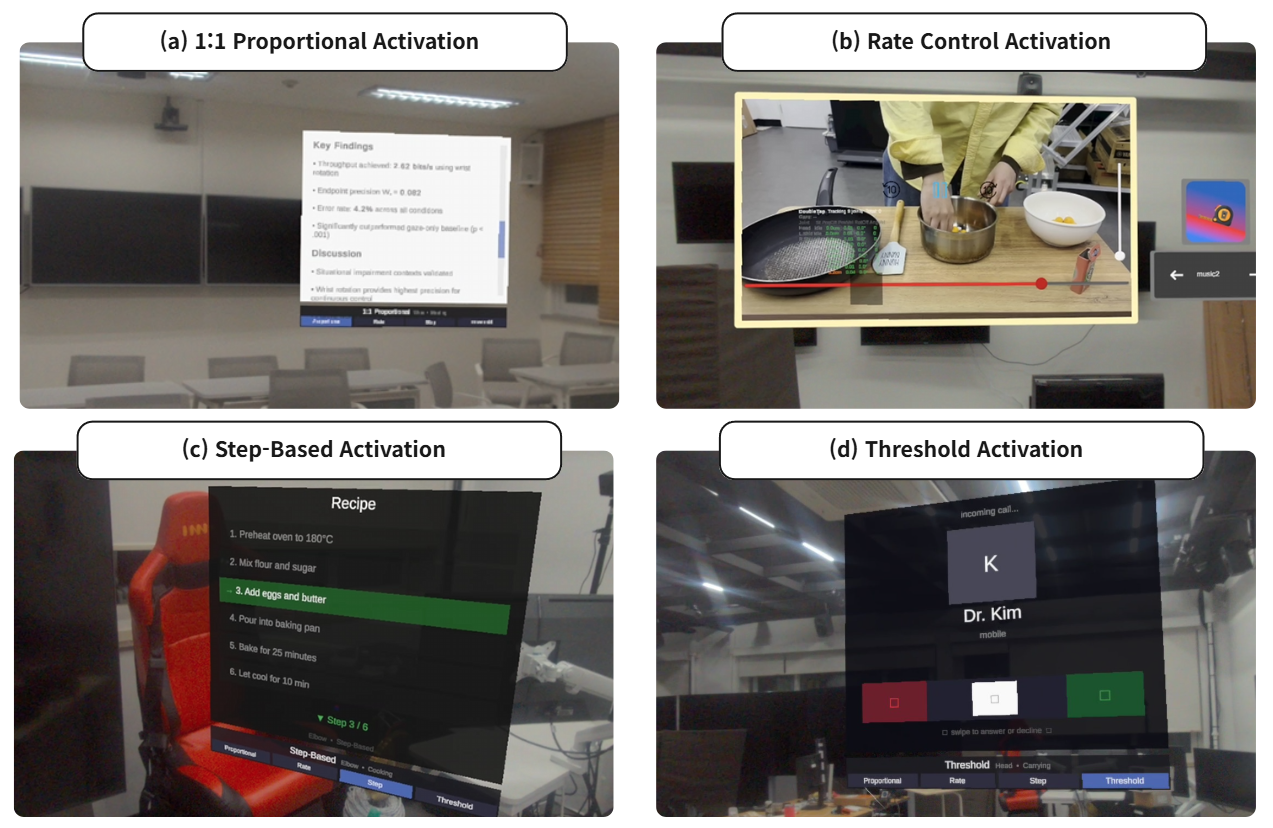}
    \caption{DOBI maps normalized spare-body movements to different UI controls: (a) 1:1 Proportional for continuous zooming, (b) Rate-Control for timeline scrubbing, (c) Step-Based for discrete navigation, and (d) Threshold Activation for binary triggers.}
    \Description{Four first-person headset screenshots of DOBI application scenarios, each with a mode selector bar along the bottom. (a) 1:1 Proportional Activation: a floating document panel titled Key Findings is displayed in a seminar room and is being continuously zoomed. (b) Rate Control Activation: a video panel showing a cooking scene is overlaid on a workshop, with a red timeline and a draggable handle being scrubbed at a controlled rate, and a music widget to the right. (c) Step-Based Activation: a dark recipe panel lists six steps with step 3, ``Add eggs and butter,'' highlighted in green and a Step 3 of 6 indicator below, showing discrete step-by-step navigation. (d) Threshold Activation: an incoming-call panel for a contact named Dr. Kim shows red decline and green accept targets, with a caption reading swipe to answer or decline, illustrating a binary trigger.}
    \label{fig:applications}
\end{figure}

DOBI normalizes ad-hoc movements into continuous 1D signals that can be filtered through various activation functions to support diverse UI behaviors without altering the underlying pipeline. We illustrate this flexibility through four scenarios spanning physical, spatial, and social constraints (Figure~\ref{fig:applications}).

\textbf{1:1 Proportional Activation: Wrist for Socially Subtle Control.} 
In professional settings where large gestures are socially conspicuous, users can perform micro-movements of the wrist under a table to scroll through AR notes or zoom into 3D models. This leverages the wrist's superior throughput (2.62~bits/s) and precision ($W_e = 0.082$) to provide highly precise, imperceptible analog control.

\textbf{Rate-Control Activation: Shoulder for Spatially Constrained Transit.} 
In crowded environments like subways, where range of motion is severely limited, users can control media timelines or volume via subtle shoulder shrugs. The shoulder's high linearity ($R^2 = 0.97$) makes it ideal for an elastic joystick metaphor; the system translates a stable posture offset into a continuous rate-controlled adjustment (e.g., fast-forwarding), requiring minimal physical space.

\textbf{Step-Based Activation: Elbow for Hands-Busy Domestic Tasks.} 
When hands are messy during tasks like cooking, the user can navigate a virtual recipe dashboard using their elbow. Given the elbow's high throughput (2.40~bits/s) and swift execution, we map vertical movements to a step-based function. This allows for deliberate, discrete jumps between recipe items or timers without the risk of overshooting.

\textbf{Threshold Activation: Head for Severe Physical Encumbrance.} 
When both hands are fully occupied (e.g., carrying a heavy box), users can manage incoming calls through discrete head tilts. By passing a predefined threshold, the movement triggers a binary "swipe" to accept or decline. This demonstrates DOBI's capacity for immediate, eyes-free binary interactions even under total limb encumbrance.

\section{Limitations and Future Work}
\label{sec:limitations}

\textbf{Scope of the current evaluation.}
Our Fitts' law study isolates DOBI's motion-to-control mapping using one pre-instructed movement per block and no concurrent physical task, establishing baseline throughput and proportionality but not evaluating free body-region switching or task continuity on its own. Our dual-task usability study (Section~\ref{sec:usability-study}) begins to address this gap, showing that DOBI remains reliable and well received when users opportunistically recruit and switch body regions during realistic primary tasks. However, this study involved only 6 participants across 3 scenarios; larger-scale, longer-duration studies are needed to confirm these trends generalize across more users, tasks, and physical contexts.

\textbf{Discrete sampling in the elicitation analysis.}
Our elicitation study sampled each articulation at five discrete control levels rather than a continuous sweep, which constrains how finely the linearity ($R^2$) of the extracted axis can be estimated and may bias it upward. The Fitts' law study offers complementary evidence from continuously produced motion ($R^2 = 0.74$--$0.97$), but a denser characterization of continuous trajectories remains future work.

\textbf{From 1D control to richer continuous input.}
This paper focuses on whether a user-recruited spare-body movement can be normalized into a robust continuous scalar. Our elicitation analysis and PCA-based mapping suggest that this 1D formulation is a useful first step, but they do not yet establish support for higher-dimensional control. Extending DOBI to 2D UI interactions such as cursor movement would require identifying two separable and user-controllable axes, rather than simply adding a second principal component. One possible direction is to estimate a trigger--conditioned control plane from repeated motion samples and pair it with activation functions designed for planar input. Evaluating whether such axes remain independent, learnable, and stable across users, joints, and contexts remains an open question.

\textbf{Scalability beyond the upper body.}
Our prototype and evaluations focus on upper-body joints, reflecting the capabilities of the current tracking setup and the design space explored in the elicitation study. The underlying pipeline from body-region recruitment, kinematic subtree extraction, to trigger-conditioned axis calibration may extend to other tracked kinematic chains, including the lower body. At the same time, lower-body interaction introduces challenges not examined here, including gait-induced motion, postural balance, and interference with locomotion. Future work should therefore investigate whether the same assumptions about trigger separability, linear control structure, and adaptive gain remain valid for knees, ankles, and feet in standing and walking contexts.

\textbf{Trigger robustness in everyday activity.}
The trigger detector was evaluated on a held-out dataset of prompted triggers and collected non-trigger motions, and the dual-task usability study further observed a low false-positive rate (1.1\%) under realistic concurrent activity. Still, data collected in a lab environment may not fully reflect real-world edge cases, including unscripted background motions, atypical articulation styles, or extended sessions beyond what we tested. Future work should evaluate trigger robustness in longer, unconstrained recordings collected outside the lab and report practical deployment metrics such as false activations over time and end-to-end interaction overhead.

\textbf{Dependence on external tracking hardware.}
Our prototype relies on external Vive Ultimate Trackers fused with the headset's built-in body tracking, which adds setup overhead and limits immediate deployability. Because the pipeline operates purely on the resulting skeletal stream rather than any tracker-specific signal, it can pair with less cumbersome pose-estimation methods to make deployment more practical: camera-based pose estimation from the headset's onboard cameras~\cite{bazarevsky2020blazepose}, sparse IMU-based full-body estimation from devices users already wear such as phones, watches, and earbuds~\cite{mollyn2023imuposer}, or neural inverse kinematics from a few body-worn sensors~\cite{jiang2024manikin}. Combined with such approaches, DOBI could run with substantially reduced or no dedicated tracking hardware; whether its trigger detection and axis extraction stay robust under the noisier, occlusion-prone estimates these methods produce remains an open question for future work.

\section{Conclusion}

We presented DOBI, a trigger-conditioned XR technique that enables users to recruit an available spare bodyregion for continuous 1D control when the hands are occupied. Through an elicitation study across six hand-busy scenarios, we found that preferred body regions vary with context, while sampled poses within a chosen articulation are often well approximated by a dominant principal axis. Building on this observation, DOBI combines gaze targeting, double-repetition trigger detection, kinematic subtree extraction, and runtime PCA-based mapping to derive a continuous control signal from a short motion sample.

A controlled 1D Fitts' law evaluation established baseline motor performance across four representative body regions, with throughputs up to 2.62 bits/s and distal joints supporting faster, more precise control than proximal ones, and a dual-task usability study showed reliable, well-received interaction (SUS = 84.2) during realistic hand-busy activities. Together, these results suggest that spare-body motion is a promising source of on-demand XR input for hand-busy interaction. More broadly, DOBI points toward XR interfaces that adapt not only to the user's task, but also to which body regions remain available for interaction at that moment.

\begin{acks}
This work was partly supported by the National Research Foundation of Korea~(NRF) grant (RS-2026-25470200, 10\%), Institute of Information \& communications Technology Planning \& Evaluation~(IITP) \& ITRC~(Information Technology Research Center) grant (IITP-2026-RS-2024-00436398, 40\%), and the IITP grant funded by the Korea government~(MSIT) (RS-2026-25507551, Development of Egocentric Data Sensing and Spatial Immersive Experience Technology, 50\%).
\end{acks}

\bibliographystyle{ACM-Reference-Format}
\bibliography{Reference}

\newpage
\appendix
\section{Dual-Task Usability Questionnaire}
\label{app:questionnaire}

Table~\ref{tab:appendix_questionnaire} lists the full text of the custom 7-point Likert items used in the dual-task usability study (Section~\ref{sec:usability-study}). Each item was rated from 1~(strongly disagree) to 7~(strongly agree), and ``[activity]'' was replaced by the participant's current primary task (cooking, carrying boxes, or walking). The six items span three concepts---Task Continuity (Q1--Q2), Control Quality (Q3--Q4), and Contextual Fit (Q5--Q6). Per-item means and standard deviations are reported in Figure~\ref{fig:usability_likert}.

\begin{table}[h]
\centering
\small
\caption{Full text of the custom 7-point Likert questionnaire items used in the dual-task usability study, grouped by concept.}
\label{tab:appendix_questionnaire}
\renewcommand{\arraystretch}{1.15}
\begin{tabularx}{\columnwidth}{@{}c l X@{}}
\toprule
\textbf{Item} & \textbf{Concept} & \textbf{Statement} \\
\midrule
Q1 & \multirow{2}{*}{Task Continuity} & ``I could operate the interface while [activity].'' \\
Q2 &                                  & ``Interacting with the interface did not break the flow of [activity].'' \\
\midrule
Q3 & \multirow{2}{*}{Control Quality} & ``I could set the slider value or operate the button as intended.'' \\
Q4 &                                  & ``The interface responded predictably to my movement.'' \\
\midrule
Q5 & \multirow{2}{*}{Contextual Fit}  & ``This technique felt natural while [activity].'' \\
Q6 &                                  & ``I could operate the interface comfortably with the hand constraints.'' \\
\bottomrule
\end{tabularx}
\end{table}

\end{document}